\documentstyle[12pt]{article}
\font\titlefont=cmbx10 scaled \magstep3

\begin{document}
\input{epsf}
\begin{flushright}
\vspace*{-2cm}
gr-qc/9702026\\ 
TUTP-96-4\\ March 20, 1997\\
\vspace*{1.5cm} 
\end{flushright}
\vskip 1in
\begin{center}
{\titlefont The unphysical nature of ``Warp Drive''}
\vskip 0.5in
Michael J. Pfenning\footnote{email: mitchel@cosmos2.phy.tufts.edu}
and L. H. Ford\footnote{email: ford@cosmos2.phy.tufts.edu}
\vskip 0.25in
{Institute of Cosmology\\ Department of Physics and
Astronomy\\ Tufts University, Medford, Massachusetts 02155, USA}
\end{center}
\vskip .6in
\begin{abstract}
We will apply the quantum inequality type restrictions to
Alcubierre's warp drive metric  on a  scale in which a local region
of spacetime can be considered ``flat''. These are inequalities 
that restrict the magnitude and extent of the negative energy
which is needed to form the warp drive metric.  From this we are able
to place limits on the parameters of the ``Warp Bubble''.  It will
be shown that the bubble wall thickness is on the order of only a few
hundred Planck lengths. Then we will show that the total integrated
energy density needed to maintain the warp metric with such thin
walls is physically unattainable.
\end{abstract}
\vskip 1in
\begin{flushleft}
Submitted to: Classical and Quantum Gravity
\end{flushleft}
\newpage


\section{Introduction}
In both the scientific community, and pop culture, humans have been
fascinated with the prospects of being able to travel between the stars
within their own lifetime. Within the framework of special relativity,
the space-going traveler may move with any velocity up to, but not
including, the speed of light.  Upon doing so, he or she would experience
a time dilation which would allow them to make the round trip from earth
to any star, and then return to earth in an arbitrarily short elapsed
time from their point of view.  However, upon returning to earth such
observers would find that their family and friends would have aged 
considerably more then they had.  This is well known as the twin 
paradox \cite{MTW, Wheeler90, Taylor}. 

Recently, Miguel Alcubierre proposed a metric  \cite{Alcu94}, fondly
called the warp drive, in which a spaceship could travel to a 
star a distance $D$ away and return home, such that the elapsed time for
the stationary observers on earth would be less than $2D/c$ where $c$ is
the velocity of light.  What is even more surprising about this spacetime
is that the proper time of the space going traveler's trip is identical
to that of the elapsed time on earth.  However, the spaceship never
{\it locally} travels faster than the speed of light.  In fact, the
spaceship can sit at rest with respect to the interior of the warp bubble.
The ship is carried along by the spacetime, much in the same way that
the galaxies are receding away from each other at extreme speeds due to
the expansion of the universe, while locally they are at rest.  The
warp drive makes use of this type of expansion (and contraction) in order
to achieve the ability to travel faster than light.

Although warp drive sounds appealing, it does have one serious drawback.
As with traversable wormholes, in order to achieve warp drive one must
employ exotic matter, that is, negative energy densities.  This is a
violation of the classical energy conditions. 
Quantum inequality restrictions in flat spacetimes on negative energies
\cite{Ford91, F&Ro92, F&Ro95, F&Ro97} do allow negative energy to exist,
however they place serious limitations on its magnitude and duration.
The flat space inequalities have been applied to the curved
spacetimes of wormhole geometries \cite{F&Ro96} with the restriction
that the negative energy be sampled on timescales
smaller than the minimum local radius of
curvature.  It was argued that over such small sampling times, the 
spacetime would be locally flat and the inequalities would be valid.
This led to the conclusion that static wormholes must either be on
the order of several Planck lengths in size, or there would be large
discrepancies in the length scales that characterize the wormhole.  

More recently, exact quantum inequalities have been developed for the
static Robertson-Walker spacetimes in three and four dimensions
\cite{Pfen96}.  In these spaces of constant curvature, it was found that
the quantum inequalities take the flat space form modified by a scale
function which depends on the ratio of the sampling time to the local
radius of curvature.  In the limit of the sampling time being smaller
than the local radius of curvature, the quantum inequalities reduce to
the flat space form, often accompanied by higher order corrections due
to the curvature \cite{Pfen96, Pfen97}. In the limit of the radius of
curvature going to infinity, one recovers the flat space inequalities
exactly.

One would like to apply the same method to the warp drive metric, but
such an exercise would require that we know the solutions to the
Klein-Gordon equation for the mode functions of the scalar field.  Such
an approach, although exact, would be exceptionally difficult.  In this
paper we will therefore apply the flat space inequality directly to the
warp drive metric
but restrict the sampling time to be small.  By doing so we will be
able to show that the walls of the warp bubble must be exceedingly
thin as compared to its radius.  This constrains the negative energy
to an exceedingly thin band surrounding the spaceship, much in the
same way it was shown that negative energy is concentrated to a thin
band around the throat of a wormhole \cite{F&Ro96}.  Recently, 
it has been shown for the Krasnikov metric \cite{Kras95}, which also
allows superluminal travel, that the required negative energy is also
constrained to a very thin wall \cite{E&Ro97}. We will then 
calculate the total negative energy that would be required to generate
a macroscopic sized bubble capable of transporting humans.  As we will
see, such a bubble would require physically unattainable energies.
 
\section{Warp Drive Basics}
Let us discuss some of the basic principles of the warp drive spacetime.
We begin with a flat (Minkowski) spacetime and then consider a small
spherical region, which we will call the bubble, inside this spacetime.
On the forward edge of the bubble, we cause spacetime to contract, and
on the trailing edge is an equal spacetime expansion.  The region inside
the bubble, which can be flat, is therefore transported forward with
respect to distant objects.  Objects at rest inside the bubble are
transported forward  with the bubble, even though they have no (or
nominal) local velocity.  Such a spacetime is described by the 
Alcubierre warp drive metric
\begin{equation}
ds^2 = -dt^2 + [dx - v_s(t)f(r_s(t))dt]^2 +dy^2+dz^2,
\label{eq:metric}
\end{equation}
where $x_s(t)$ is the trajectory of the center of the bubble and
$v_x(t)= dx_s(t) / {dt}$ is the bubble's velocity. The variable 
$r_s(t)$ measures the distance outward from the center of the bubble
given by
\begin{equation}
r_s(t) = \sqrt{(x-x_s(t))^2 +y^2+z^2}.
\end{equation}
The shape function of the bubble is given by $f(r_s)$, which
Alcubierre originally chose to be
\begin{equation}
f(r_s) = {{\tanh[\sigma(r_s +R)] - \tanh[\sigma(r_s -R)]}\over 
2 \tanh[\sigma\; R]}. \label{eq:alcub}
\end{equation}
The variable $R$ is the radius of the warp bubble, and $\sigma$ is a 
free parameter which can be used to describe the thickness of the
bubble walls. In the large $\sigma$ limit, the function $f(r_s)$ quickly
approaches that of a top hat function, where $f(r_s) = 1$ for $r_s 
\leq R$ and zero everywhere else.  It is not necessary to choose
a particular form of $f(r_s)$. Any function will suffice so long as it
has the value of approximately 1 inside some region of $r_s<R$ and
goes to zero rapidly outside the bubble, such that as $r_s \rightarrow 
\infty$ we recover Minkowski space. In order to make later calculations
easier, we will also use the piece-wise continuous function
\begin{equation}
f_{p.c.}(r_s) = \left\{\matrix{1 & r_s < R-{\Delta\over2}\cr
-{1\over\Delta}(r_s -R-{\Delta\over2})\qquad & R-{\Delta\over2}<r_s<
R+{\Delta\over2}\cr
0&r_s > R+{\Delta\over2}}\right.
\label{eq:f_approx}
\end{equation}
where $R$ is the radius of the bubble.  The variable $\Delta$ is the
bubble wall thickness.  It is chosen to relate to the parameter
$\sigma$ for the Alcubierre form of the shape function by setting the
slopes of the functions $f(r_s)$ and $f_{p.c.}(r_s)$ to be equal at
$r_s=R$. This leads to
\begin{equation}
\Delta = {\left[1+\tanh^2(\sigma R)\right]^2\over
 2\;\sigma\;\tanh(\sigma R)}\;,
\end{equation}
which in the limit of large $\sigma R$ can be approximated
by $\Delta \simeq 2/\sigma$.  

We now turn our attention to the solutions of the geodesic equation.
It is straightforward to show that  
\begin{equation}
{dx^\mu \over dt }= u^\mu = \left( 1, v_s(t)f(r_s(t)),0,0 \right),\qquad
u_\mu = (-1,0,0,0)
\end{equation}
is a first integral of the geodesic equations.  Observers with this
four-velocity are called the Eulerian observers by Alcubierre.  We see
that the proper time and the coordinate time are the same for all 
observers. Also, the y and z components of the 4-velocity are zero.
The bubble therefore exerts no ``force'' in the directions perpendicular
to the direction of travel. In Figure 1, we have plotted one such
trajectory for a observer that passes through the wall of a warp bubble
at a distance $\rho$ away from the center of the bubble. The x-component
of the 4-velocity is dependent on the shape function, and solving this
explicitly for all cases can be rather difficult due to the time
dependence of $r_s(t)$. A spacetime plot of an observer with the
four-velocity given above is shown in Figure 2, for a bubble with
constant velocity.
\begin{figure}
\begin{center}\leavevmode\epsfysize=6cm\epsffile{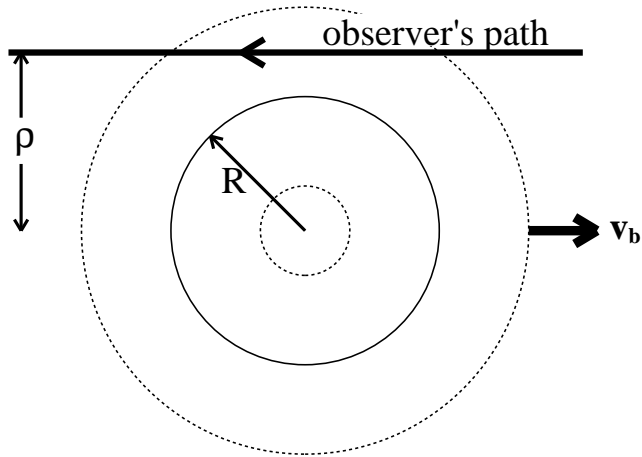}\end{center}
\caption{The path of an observer who passes through the outer region
of the bubble, shown in the bubble's rest frame.  As viewed from the 
interior of the bubble, the observer is moving to the left.}
\end{figure}
\begin{figure}
\begin{center}\leavevmode\epsfbox{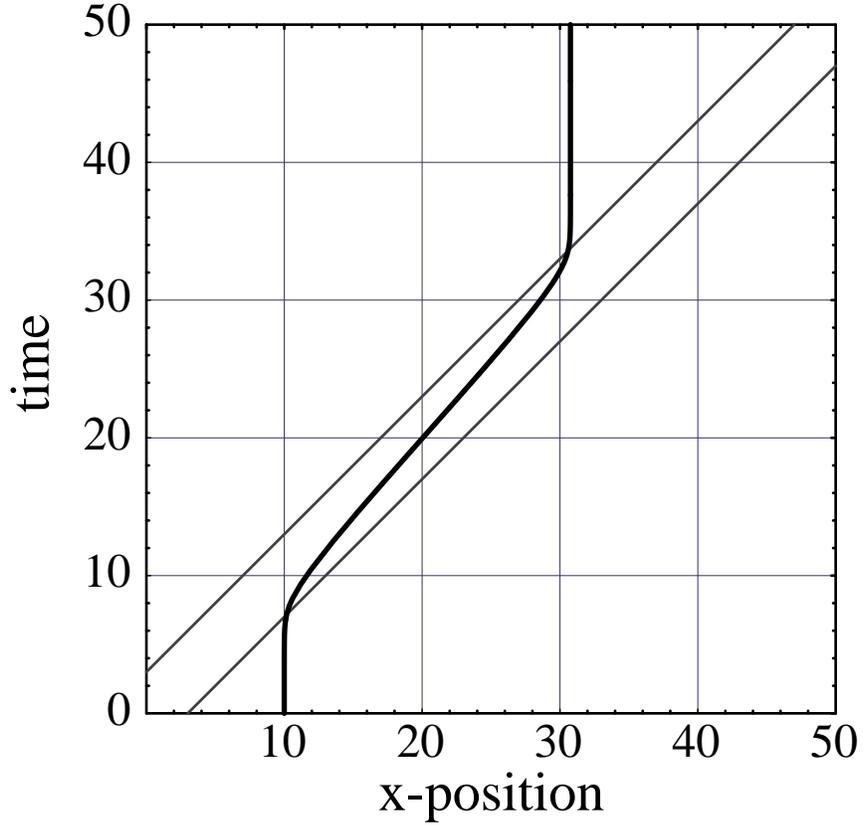}\end{center}
\caption{The worldline (the dark line) of the geodesic observer
passing through the outer region of a warp bubble, plotted in the
observer's initial rest frame.  The two lighter diagonal lines are
the worldlines of the center of the bubble wall on the front and
rear edges of the bubble, respectively. The bubble has a radius
of 3, a velocity of 1, and the $\sigma$ parameter is also 1. The
plot shows an observer who begins at rest at $x=10$, $y^2+z^2 =
\rho^2 = 4$.  The shape function is of the form given by Alcubierre,
Equation~(\ref{eq:alcub}).}
\end{figure}

We see that the Eulerian observers are initially at rest.  As the
front wall of the bubble reaches the observer, he or she begins to
accelerate, relative to observers at large distances, in the
direction of the bubble.  Once inside the bubble
the observer moves with a nearly constant velocity given by
\begin{equation}
{dx(t)\over dt}|_{max.} = v_s(t_\rho) f(\rho),
\label{eq:Vmax}
\end{equation}
which will always be less than the bubble's velocity unless $\rho 
= (y^2+z^2)^{1/2} = 0$. The time $t_\rho$ is defined by $r_s(t_\rho)
= \rho$, i.e. it is the time at which the observer reaches the bubble 
equator.  Such observers then decelerate, and are left at rest as
they pass out of the rear edge of the bubble wall.  In other words no
residual momentum is imparted to these observers during the ``collision''.
However they have been displaced forward in space along the trajectory
of the bubble.

There is also another interesting feature of these geodesics.
As already noted, the observers will move with a nearly constant
velocity through the interior of the bubble. This holds true for
any value of $\rho$.  However, the velocity is still dependent
upon the value of $\rho$, so observers at different distances 
from the center of the bubble will be moving with different
velocities relative to one another. 
If a spaceship of finite size is placed inside the bubble
with its center of mass coincident with the center
of the bubble, then the ship would experience a net ``force'' pushing it
opposite to the direction of motion of the bubble, so long $1-f(r_s)$
is nonzero at the walls of the ship. 
The ship would therefore have to use its engines to maintain its
position inside the bubble.  In addition, the ship would be subject
to internal stresses on any parts that extended sufficiently far away  
from the rest of the ship.

In the above discussion we have used the Alcubierre form of the shape
function, $f(r_s)$.  If one uses the piece-wise continuous form,
Equation~(\ref{eq:f_approx}), one
finds similar results with some modification.  Inside the bubble, where
$r_s < (R-{\Delta/2})$, every observer would move at exactly the speed
of the bubble.  So any observer who reaches the bubble interior would
continue on with it forever.  This arises from the fact that everywhere
inside the bubble, spacetime is perfectly flat because $f(r_s) = 1$.
For observers whose geodesics pass solely through the bubble walls,
so $(R-{\Delta/2}) < \rho < (R+{\Delta/2})$, the result is more or
less identical to that of the geodesics found with the Alcubierre shape
function.  This is the region we are most interested in because it is
the region that contains the largest magnitude of negative energy.
 
We now turn our attention to the energy density distribution of the
warp drive metric.  Using the first integral of the geodesic equations,
it is easily shown that
\begin{equation}
\langle T^{\mu\nu}u_\mu u_\nu\rangle 
 = \langle T^{00} \rangle = {1\over 8\pi}G^{00} = -{1\over 8\pi}
{v_s^2(t) \rho^2 \over 4 r_s^2(t) }\left(d f(r_s)\over dr_s
\right)^2,
\label{eq:EnergyDensity}
\end{equation}
where $\rho = [y^2 + z^2]^{1/2}$, is the radial distance perpendicular
the $x$-axis as was defined above. We immediately see that the energy
density measured by any geodesic observer is always negative, as was
shown in Alcubierre's original paper\cite{Alcu94}.  In Figure 3, we
see that the distribution of negative energy is concentrated in a
toroidal region perpendicular to the direction of travel.
\begin{figure}
\begin{center}\leavevmode\epsfysize=8.55cm\epsffile{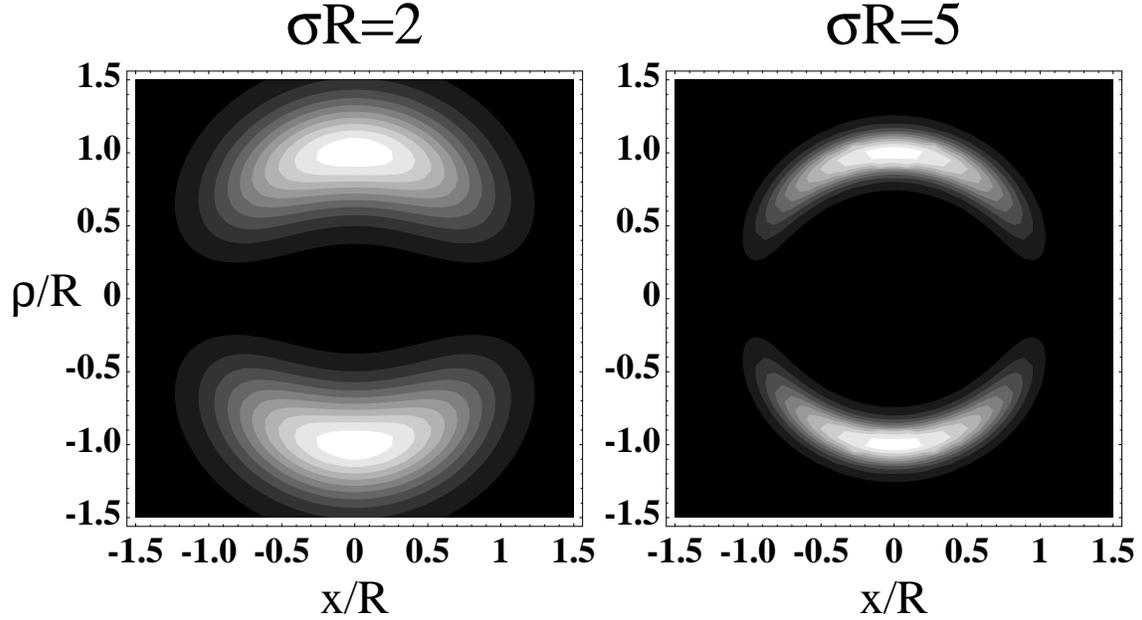}\end{center}
\caption{The negative energy density is plotted for a longitudinal
cross section of the warp metric traveling at constant velocity
$v_s = 1$ to the right for the Alcubierre shape function.  Black
regions are devoid of matter, while white regions are maximal
negative energy.}
\end{figure}

In Section 4 we will integrate the energy density over all of space
to obtain the total negative energy required to maintain the bubble,
under the restrictions of the quantum inequalities.  As we will show,
the total energy is physically unrealizable in the most extreme sense.  
 
\section{Quantum Inequality Restrictions}
We begin with the quantum inequality (QI) for a free, massless
scalar field in four-dimensional Minkowski spacetime derived by Ford
and Roman \cite{F&Ro95},  
\begin{equation}
{\tau_0\over\pi}\int_{-\infty}^\infty {{\langle T_{\mu\nu}u^\mu u^\nu
 \rangle}\over{\tau^2 + \tau_0^2}}d\tau\geq -{3\over{32\pi^2 \tau_0^4}}
 \; ,\label{eq:original}
\end{equation}
where $\tau$ is an inertial observer's proper time, and $\tau_0$ 
is an arbitrary sampling time.
This places a limit on the magnitude and duration of the negative
energy density experienced by an observer.  In the limit that
$\tau_0 \rightarrow \infty$ one recovers the Averaged Weak Energy
Condition (AWEC).  It has been argued by Ford and Roman \cite{F&Ro96}
that one may apply the QI to non-Minkowski spacetimes if the sampling
time is of the order, or less than the smallest
local radius of curvature. 

We begin by taking the expression for the energy density (\ref
{eq:EnergyDensity}), and inserting it into the quantum inequality,
Equation~(\ref{eq:original}).  One finds
\begin{equation}
t_0 \int_{-\infty}^{+\infty} {v_s(t)^2 \over r_s^2} \left({df(r_s)\over
dr_s}\right)^2 {dt\over {t^2+t_0^2}} \leq {3\over \rho^2 t_0^4} \, .
\label{eq:explt_form}
\end{equation}
If the time scale of the sampling is sufficiently small compared
to the time scale over which the bubble's velocity is changing, then
the warp bubble's velocity can be considered roughly constant,
$v_s(t) \approx v_b$, during the sampling interval. 
We can now find the form of the geodesic at the time the sampling is
taking place. Because of the small sampling time, the
$[t^2 +t_0^2]^{-1}$ term becomes strongly peaked, causing the QI
integral to sample only a small portion of the geodesic.  We therefore
arrange that the observer is at the equator of the warp bubble at
$t = 0$. Then the geodesic is well-approximated by
\begin{equation}
x(t) \approx f(\rho) v_b t \; ,
\end{equation}
which results in
\begin{equation}
r_s(t) = \left[ (v_b t)^2 (f(\rho) - 1)^2 + \rho^2 \right]^{1/2}\;.
\end{equation}

Finally, we must specify the form of the shape function of the bubble.
If we Taylor series expand any shape function about the sampling point,
$r_s(t) \rightarrow \rho$, and then take the appropriate
derivatives to obtain the needed term for the quantum inequality,
we find
\begin{equation}
{df(r_s)\over dr_s} \approx f'(\rho) + f''(\rho)[r_s(t) - \rho]+....
\end{equation}
The leading term is the slope of the shape function at the sampling
point, which is in general roughly proportional to the  inverse of
the bubble wall thickness.  We can therefore use, with no loss of
generality, the piece-wise continuous form of the shape function 
(\ref{eq:f_approx}) to obtain a good order of magnitude 
approximation for any choice of shape function.
The quantum inequality (\ref{eq:explt_form}) then becomes
\begin{equation}
 t_0 \int_{-\infty}^{+\infty} {dt \over (t^2 + \beta^2) (t^2+t_0^2)} 
\leq {3 \Delta^2 \over v_b^2 \, t_0^4 \, \beta^2}
\end{equation}
where
\begin{equation}
\beta = {\rho \over {v_b (1 - f(\rho) ) }}\, .
\end{equation}
Formally the integral should not be taken over all time but just the
time the observer is inside the bubble walls.  However, the sampling 
function rapidly approaches zero.  Therefore contributions to the
integral from the distant past or the far future are negligible.  The
integral itself can be done as the principal value of a contour that
is closed in the upper half of the complex plane.  We find 
\begin{equation}
\int_{-\infty}^{+\infty} {dt \over (t^2 + \beta^2) (t^2+t_0^2)}
 = {\pi \over t_0 \;\beta\; (t_0 + \beta)},
\end{equation}
yielding an inequality of
\begin{equation}
{\pi \over 3} \leq {\Delta^2 \over {v_b^2 \; t_0^4}} 
\left[{v_b t_0 \over \rho} (1-f(\rho)) +1\right] \; .
\end{equation}

The above inequality is only valid for sampling times on which the
spacetime may be considered approximately flat.  We must therefore
find some characteristic length scale below which this occurs.  For
an observer passing through the bubble wall at a distance $\rho$ from
the center, one may calculate the Riemann tensor in the static 
background frame, then transform the components to the observer's
frame by use of an orthonormal tetrad of unit vectors.  In this
frame, the tetrad is
given by the velocity vector $u^\mu (t)$ and three unit vectors
$\hat x$, $\hat y$, and $\hat z$.  One finds that the largest component 
of the Riemann tensor in the orthonormal frame is given by
\begin{equation}
|R_{{\hat t}{\hat y}{\hat t}{\hat y}}| = 
{3 v_b^2\; y^2 \over 4\;\rho^2}\left[ {d f(\rho)
\over d\rho} \right]^2
\end{equation}
which yields
\begin{equation}
r_{min} \equiv {1\over \sqrt{|R_{{\hat t}{\hat y}{\hat t}{\hat y}}|}}
\sim {2\Delta \over {\sqrt 3}\; v_b}\; ,
\end{equation}
when $y= \rho$ and the piece-wise continuous form of
the shape function is used.  The sampling time must be smaller than
this length scale, so we take
\begin{equation}
t_0 = \alpha {2\Delta \over {\sqrt 3}\; v_b} \qquad 0<\alpha\ll 1 .
\end{equation}
Here $\alpha$ is an unspecified parameter that describes how much
smaller the sampling time is compared to the minimal radius of
curvature. If we insert this into the quantum inequality and use
\begin{equation}
{\Delta \over \rho} \sim {v_b t_0 \over \rho} \ll 1 \; ,
\end{equation}
we may neglect the term involving $1-f(\rho)$ to find
\begin{equation}
\Delta \leq {3\over 4}\sqrt{3\over\pi}\;{v_b \over \alpha^2}\, .
\end{equation}
Now as an example, if we let $\alpha = 1/10$, then
\begin{equation}
\Delta \leq 10^2\, v_b\; L_{Planck}\, ,
\label{eq:wall_thickness}
\end{equation}
where $L_{Planck}$ is the Planck length.  Thus, unless $v_b$ is
extremely large, the wall thickness cannot be much above the Planck
scale. Typically, the walls of the warp bubble are so thin that the
shape function could be considered a ``top hat'' for most purposes.

\section{Total Energy Calculation}
We will now look at the total amount of negative energy that is
involved in the maintenance of a warp metric.  For simplicity, let us
take a bubble that moves with constant velocity such that $x_s(t)
= v_b \;t$. Because the total energy is constant, we can calculate it
at time $t=0$. We then have
\begin{equation}
r_s(t=0) = [x^2+y^2+z^2]^{1\over 2} = r.
\end{equation}
With this in mind we can write the integral of the local matter energy
density over proper volume as
\begin{equation}
E = \int dx^3 \sqrt{|g|}\;\, \langle T^{00}\rangle = -{v_b^2 \over 32\pi}
\int {\rho^2 \over r^2} \left(d f(r)\over dr \right)^2 dx^3 \; ,
\end{equation}
where $g = {\rm Det}|g_{ij}|$ is the determinant of the spatial metric
on the constant time hypersurfaces. Portions of this integration can be
carried out by making a transformation to spherical coordinates.  By
doing so, one finds that
\begin{equation}
E = - {1\over 12} v_b^2 \int_0^\infty r^2
\left( d\;f(r)\over dr\right)^2 dr
\end{equation}
Since we are making only order of magnitude estimates of the
total energy, we will use a piece-wise continuous approximation
to the shape function given by Equation~(\ref{eq:f_approx}). When one
takes the derivative of this shape function, we find that the
contributions to the energy come only from the bubble wall region,
and we end up evaluating
\begin{eqnarray}
E &=&- {1\over 12} v_b^2 \int_{R-{\Delta\over 2}}^{R+{\Delta\over 2}} 
r^2 \left(- 1 \over \Delta\right)^2 dr\\
&=&- {1\over 12} v_b^2\left( {R^2 \over\Delta}+{\Delta\over 12}\right).
\label{eq:energy}
\end{eqnarray}
For a macroscopically useful warp drive, we want the radius of the
bubble to be at least in the range of 100 meters so that we may
fit a ship inside.  It has been shown in the previous section that
the wall thickness is constrained by (\ref{eq:wall_thickness}).
If we use this constraint and let the bubble radius be equal to
100 meters, then we may neglect the second term on the right-hand-side
of Equation~(\ref{eq:energy}).  
It follows that
\begin{equation}
E \leq -6.2\times10^{70}\, v_b \; L_{Planck} \; \sim\;
-6.2\times10^{65}\, v_b \; {\rm grams}.
\end{equation}
Because a typical galaxy has a mass of approximately
\begin{equation}
M_{Milky Way} \approx 10^{12}\; M_{sun} = 2\times 10^{45} {\rm grams},
\end{equation}
the energy required for a warp bubble is on the order of 
\begin{equation}
E \leq - 3 \times 10^{20} \; M_{galaxy} \; v_b\; .
\end{equation}
This is a fantastic amount of negative energy, roughly ten orders of
magnitude greater than the total mass of the entire visible universe.

If one can violate the quantum inequality restrictions and make
a bubble with a wall thickness on the order of a meter, things are
improved somewhat.  The total energy required in the case of the
same sized radius and $\Delta = 1$ meter would be on the order
of a quarter of a solar mass, which would be more practical, yet
still not attainable.

\section{Summary}
We see that, from (\ref{eq:wall_thickness}), the quantum
inequality restrictions on the warp drive metric constrain the
bubble walls to be exceptionally thin.  Typically, the walls are on
the order of only hundreds or thousands of Planck lengths.
Similar constraints on the size of the negative energy region
have been found in the case of traversable wormholes \cite{F&Ro96}.

One might note that by making the velocity of the bubble, $v_b$,
very large then we can make the walls thicker, however this
causes another problem.  For every order of magnitude by which
the velocity increases, the total negative energy required to
generate the warp drive metric also increases by the same magnitude.
It is evident, that for macroscopically sized bubbles to be useful
for human transportation, even at subluminal speeds, the required
negative energy is unphysically large.  

On the other hand, we may consider the opposite regime.  Warp bubbles
are still conceivable if they are very tiny, i.e., much less than the
size of an atom.  Here the difference in length scales is not as great.
As a result, a smaller amount of negative energy is required to maintain
the warp bubble.  For example, a bubble with a radius the size of 
one electron Compton wavelength would require
a negative energy of the order $E \sim - 400 M_{sun}$.

The above derivation assumed that we are using a quantized, massless
scalar field to generate the required negative energy.  Similar
quantum inequalities have been proven for both massive scalar fields
\cite{F&Ro97,Pfen96} and the electromagnetic field \cite{F&Ro97}.
In the case of the massive scalar field, the quantum inequality
becomes even more restrictive, thereby requiring the bubble walls
to be even thinner.  For the quantized electromagnetic field, the
wall thickness can be made larger by a factor of $\sqrt{2}$, due
to the two spin degrees of freedom of the photon.  However this is
not much of an improvement over the scalar field case.

\centerline {\bf Acknowledgments}

{We would like to thank Thomas A. Roman and Allen Everett for
useful discussion. This research was supported in part by NSF Grant
No.~Phy-9507351 and the John F. Burlingame Physics Fellowship Fund.}

\end{document}